\documentclass[runningheads]{llncs}
\usepackage[T1]{fontenc}
\usepackage{graphicx}
\usepackage{amsmath}
\usepackage{graphicx}
\usepackage{array}
\usepackage{adjustbox}
\usepackage{tabularx}
\usepackage{booktabs}

\usepackage{hyperref}

\hypersetup{
    colorlinks=true,
    linkcolor=blue,
    citecolor=blue,
    urlcolor=blue
}

\begin{document}
%
% \title{Automatic Brain Tissue Segmentation from MRI with N4BiasFieldCorrection and Anisotropic Diffusion Pre-processing Techniques: A Comparative Study between Statistical and Deep Learning Approaches}
\title{Comparative Study of Probabilistic Atlas and Deep Learning Approaches for Automatic Brain Tissue Segmentation from MRI Using N4 Bias Field Correction and Anisotropic Diffusion Pre-processing Techniques}
\titlerunning{Probabilistic Atlas and Deep Learning for Brain MRI Segmentation}
% If the paper title is too long for the running head, you can set
% an abbreviated paper title here
%
\author{Mohammad Imran Hossain\inst{1,2,3},
Muhammad Zain Amin \inst{1,2,3}, Daniel Tweneboah Anyimadu\inst{1,2,3} \and Taofik Ahmed Suleiman\inst{1,2,3,4}}
\authorrunning{Mohammad Imran Hossain et al.}
% First names are abbreviated in the running head.
% If there are more than two authors, 'et al.' is used.
%
%%%%% New %%%%%
\institute{Department of Medical Imaging and Computing, University of Girona, Girona, Spain
\\
\and
Graduate School of Science and Technology,  University of Burgundy, Le Creusot, France
\\
\and
Department of Electrical and Information Engineering, University of Cassino and Southern Lazio, Cassino, Italy
\\
\and
Wallace H. Coulter Department of Biomedical Engineering, Georgia Tech and Emory University, Atlanta, GA, USA.}

%%%%% New %%%%%
% \institute{Universitat de Girona, Spain \\ 
% \email{u1984911@campus.udg.edu}\\
% \and
% Carl E. Ravin Advanced Imaging Laboratories, Duke University School of Medicine, Durham, NC, USA \\
% \email{muhammadzain.amin@duke.edu}\\
% \and
% Wallace H. Coulter Department of Biomedical Engineering, Georgia Tech and Emory University, Atlanta, GA, USA \\
% \email{tsuleiman6@gatech.edu}}
%
\maketitle              % typeset the header of the contribution
\begin{abstract}
Automatic brain tissue segmentation from Magnetic Resonance Imaging (MRI) images is vital for accurate diagnosis and further analysis in medical imaging. Despite advancements in segmentation techniques, a comprehensive comparison between traditional statistical methods and modern deep learning approaches using pre-processing techniques like N4 Bias Field Correction and Anisotropic Diffusion remains underexplored. This study provides a comparative analysis of various segmentation models, including Probabilistic ATLAS, U-Net, nnU-Net, and LinkNet, enhanced with these pre-processing techniques to segment brain tissues (white matter (WM), grey matter (GM) and cerebrospinal fluid (CSF)) on the Internet Brain Segmentation Repository (IBSR18) dataset. Our results demonstrate that the 3D nnU-Net model outperforms others, achieving the highest mean Dice Coefficient score (0.937 ± 0.012), while the 2D nnU-Net model recorded the lowest mean Hausdorff Distance (5.005 ± 0.343 mm) and the lowest mean Absolute Volumetric Difference (3.695 ± 2.931 mm) across five unseen test samples. The findings highlight the superiority of nnU-Net models in brain tissue segmentation, particularly when combined with N4 Bias Field Correction and Anisotropic Diffusion pre-processing techniques. Our implemented code can be accessed via GitHub \href{https://github.com/imran-maia/IBSR_18_BraTSeg_Deep_Learning}{\textit{link}}.

\end{abstract}
%
%
%

%===================================================================
\section{Introduction}
Image segmentation divides images into regions with homogeneous characteristics and is vital for analyzing medical structures \cite{dora2017state,serra2003image,sharma2010automated}. In brain MRI, segmenting tissues like White Matter (WM), Gray Matter (GM), and Cerebrospinal Fluid (CSF) is essential for diagnosing neurological conditions such as brain tumors, multiple sclerosis, and traumatic brain injuries \cite{despotovic2015mri}. The effectiveness of brain tissue segmentation methods is highly influenced by factors including location, size, shape, texture, and contrast, particularly when using different imaging sequences \cite{wen2015novel}.

Recent advancements in brain tissue segmentation from MRI have explored both statistical and deep learning approaches. Conventional techniques, including region-based methods \cite{gould2009region} that group pixels by color, intensity, or texture; clustering methods \cite{mittal2022comprehensive} that group pixels based on similarity or proximity; statistical methods like Gaussian mixture models \cite{riaz2020gaussian} and atlas-based techniques\cite{cabezas2011review}; and classification methods using engineered features with machine learning classifiers, have been widely used due to their simplicity and interpretability \cite{dora2017state}. However, these methods often suffer from sequential processing errors, leading to suboptimal results.

On the other hand, Convolutional Neural Networks (CNNs) have revolutionized brain tissue segmentation by providing end-to-end solutions that automatically learn relevant features from MRI data \cite{havaei2017brain,tushar2019brain}. CNN-based models like U-Net \cite{ronneberger2015u}, which utilizes skip connections to preserve spatial information, and LinkNet \cite{chaurasia2017linknet}, a lightweight architecture with efficient encoder-decoder paths, have demonstrated significant improvements in segmentation accuracy and computational efficiency.

Notable studies on brain tissue segmentation from MRI include \cite{makropoulos2014automatic}, who achieved high Dice Similarity Coefficient (DSC) performance with an atlas-based method, and \cite{lee2020automatic}, who reported an average DSC of 0.93 using a Patch-wise U-Net for segmenting Open Access Series of Imaging Studies (OASIS) and Internet Brain Segmentation Repository (IBSR) brain tissues. Additionally, \cite{baniasadi2023dbsegment} used CNNs from the nnU-Net \cite{isensee2021nnUNet} framework to segment T1 images from multiple datasets, obtaining an average DSC of 0.89 ± 0.04, while \cite{ramzan2020volumetric} utilized 3D CNNs with residual learning to achieve DSCs of 0.879 and 0.914 for segmenting brain tissues into 3 and 9 regions, respectively.

Despite significant advancements in brain tissue segmentation, a comprehensive comparison between traditional statistical methods and modern deep learning approaches, particularly with pre-processing techniques like N4 Bias Field Correction \cite{tustison2010n4itk} and Anisotropic Diffusion, is lacking. This study aims to fill this gap by systematically evaluating these methods on the IBSR18 dataset. We explore the effectiveness of various segmentation models, including Probabilistic ATLAS, U-Net, nnU-Net, and LinkNet, using encoder backbones like ResNet34 and ResNet50. Evaluation metrics such as Dice Coefficient Score (DSC), Hausdorff Distances (HD), and Absolute Volumetric Differences (AVD) are used to assess performance. The paper is structured further to discuss methodologies, present experiments and results, and analyze the findings’ implications in medical imaging.

%======================================================================

\section{Methodology}

%====================================================================

\subsection{Dataset}
In this study, we used the IBSR18 dataset, a publicly available dataset provided by the Center for Morphometric Analysis at Massachusetts General Hospital, USA \cite{ibsr18nitrc}. The dataset consists of 18 T1-weighted volumes with varying slice thickness and spacing. Out of 18 samples, the dataset was divided into three subsets for the segmentation task: eight volumes for training, two volumes for validation, and five for testing. The training subset was used to train the models, the validation subset was employed to evaluate and fine-tune model performance, and the test subset was utilized to make the final predictions. 
%====================================================================

% ================================================================

\subsection{Preprocessing}
 Segmenting brain tissue from MRI presents unique challenges due to variations in scanner settings. These differences often result in uneven intensity levels, contrast variations, and assorted noise types. As a result, getting uniform and precise tissue segmentation becomes difficult. Preprocessing steps are crucial to address these issues and attain optimal segmentation results. This preprocessing typically includes skull-stripping, bias field correction, and image homogenization. As the volumes of IBSR18 are already skull-stripped, we did not perform any skull-stripping. However, we corrected the bias field using N4 bias field correction \cite{tustison2010n4itk} and further denoised the intensities of all the volumes using anisotropic diffusion. The pipeline for preprocessing is shown in Figure 1.

\begin{figure}
\includegraphics[width=\textwidth]{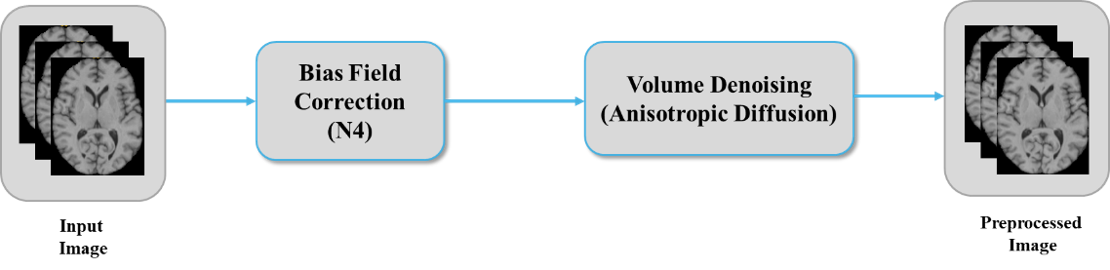}
\caption{ Pipeline for MR image preprocessing.
} \label{fig1}
\end{figure}

%=====================================================================

%=====================================================================

\subsection{Patch Extraction}

Training deep learning models for segmentation with higher dimensional images requires strong computational resources and time, which is always challenging to manage. However, one of the widely used techniques to overcome the computational issue is to extract small patches from the original higher dimensional images and later use them to train deep learning models. In our study, we employed deep learning architectures like U-Net \cite{ronneberger2015u}, nnU-Net \cite{isensee2021nnUNet}, and LinkNet\cite{chaurasia2017linknet}, and to reduce the computational complexity during training, we extracted patches of size 32×32 from the full 3D volumes along with their corresponding ground truths from the training subset. This approach not only streamlined our training process but also ensured the effective utilization of computational resources by excluding the backgrounds while maintaining the quality of the segmentation results.

To achieve this step, we started by defining the size of the patches and the stride size of 32×32. We then created a mask to isolate the brain tissue regions from the background in the volumes, focusing exclusively on the tissue regions and finally extracting the patches as shown in Figure 2.

\begin{figure}
\includegraphics[width=\textwidth]{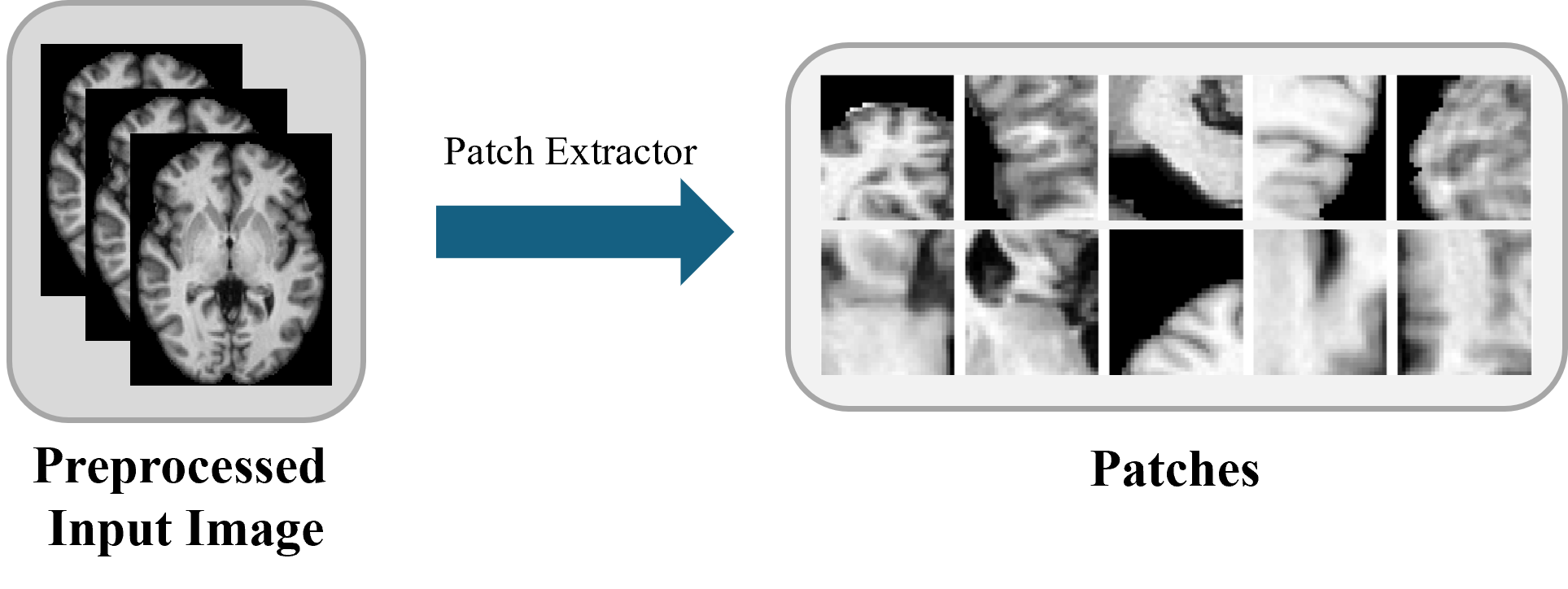}
\caption{ Patch extraction.
} \label{fig1}
\end{figure}

%=====================================================================

%=====================================================================

\subsection{Data Augmentation}
Data augmentation is an important process in deep learning that artificially expands the training subset by applying various transformations to the existing images. This process generates new training examples that simulate potential variations in real-world data, thereby reducing the risk of overfitting, and improving the model's robustness and performance on unseen data. In this study, we focused on optimizing the training of the U-Net and LinkNet models, using the training subset consisting of eight volumes.  We allocated two volumes for validation, adhering to a standard training-validation split. 

The implemented data augmentation strategy includes a range of transformations such as rotations, shifts in width and height, shear transformations, scaling, and both horizontal and vertical flipping. With these diverse augmentations, we created an enriched training dataset that provided the model with a wide array of image conditions. This approach is particularly beneficial in medical image analysis, such as brain tissue segmentation, where the model needs to be robust enough to handle various image orientations and anomalies. 
%=====================================================================

%=====================================================================

\subsection{Brain Tissue Segmentation Using Probabilistic ATLAS}

An atlas, particularly in the context of brain tissue segmentation from MR images, is a crucial tool in medical imaging and computational anatomy. It is a combination of two elements: an intensity image, often referred to as the template, and its corresponding segmented image containing atlas labels. To segment images, they are initially aligned or registered to this template. Following the registration, the obtained transformation is then applied in reverse to the atlas labels. This reverse application leads to the creation of the segmentation outcome for the target image, constituting the entire process known as label propagation. The approach used for label propagation can vary based on the type of atlas utilized. The atlases can be categorized into two primary categories such as topological atlas and probabilistic atlas.

For our study, we implemented a probabilistic atlas after aligning all the images in the dataset to a common reference image and transferring their labels to this shared registration space. The choice of the reference image during the alignment process was based on computing the mutual information amongst the training images. The transferred labels can then be merged using either the averaging technique i.e. calculating the average of the transferred labels for each class or the majority voting where each voxel is assigned the label that appears most. The process of constructing our atlas typically involves three key steps:

\begin{enumerate}
    \item  \textit{Image Registration:} This is done to align the tissue region voxels of all the images in training data. To register the images, we chose a fixed image based on the mutual information, and all other images in the training set were considered moving images and registered against the fixed image to align them into a common space. Several tools and approaches can be used for performing image registration, however, we used the SimpleITK Elastix for our registration using different registration parameters such as Rigid, Affine, and B-Spline to better fine-tune and get the best results of the brain image registration.
    
    \item \textit{Label Propagation:} After the registration, the subsequent task is to propagate the label. In this step, the transformation parameters acquired in registering the images are applied to the corresponding labels for each instance. This process ensures that labels are moved to the registration space the same way as their images. For this step, we used the SimpleITK Transformix tool.
    
    \item \textit{ATLAS Generation:} After the image registration and corresponding label propagation is completed for each sample in the training subset, we then generate the atlas. The choice of the atlas template is pivotal and for optimal results, we used the mean image derived from all the registered images. This mean image is calculated voxel-wise across the dataset, using the formula:

    \begin{equation}
    I(x, y, z)_M = \frac{1}{N} \sum_{i=1}^{N} I(x, y, z)
    \end{equation}

    Where, \( I(x, y, z)_M \) represents the value of the voxel at position \((x, y, z)\) in the mean image, \( N \) is the total number of images in the dataset, and \( I_i(x, y, z) \) denotes the value of the voxel at position \((x, y, z)\) in the \(i\)-th registered image. For label synthesis, the propagated labels from the training images are combined using majority voting. These approaches ensure accurate and representative labeling in the atlas, reflecting the most common features across the dataset. Figure 3 shows the full pipeline for our brain tissue segmentation using a probabilistic atlas.
\end{enumerate}

\begin{figure}
\includegraphics[width=\textwidth]{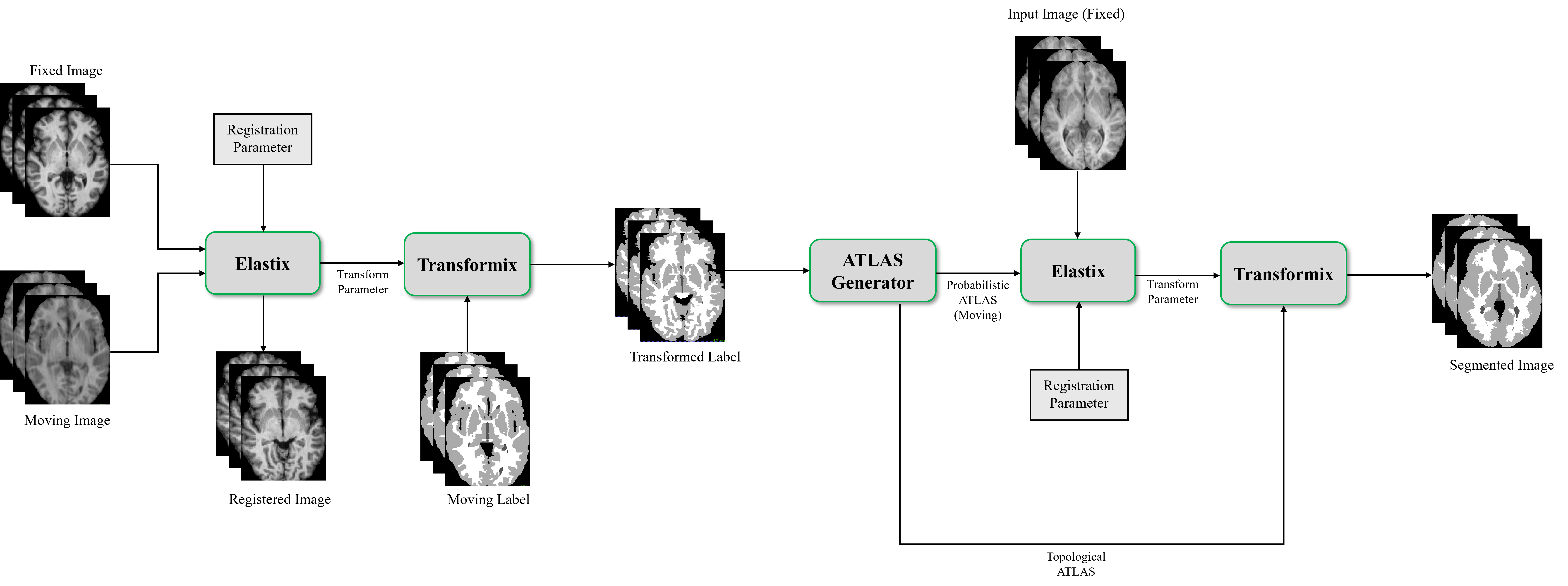}
\caption{Pipeline for brain tissue segmentation using probabilistic atlas.
} \label{fig1}
\end{figure}

\subsection{Brain Tissue Segmentation Using Deep Learning}

Deep learning leverages complex neural network architectures to learn from large datasets, and since we have a limited amount of data, we initially implemented a statistical approach based on a probabilistic atlas to serve as our baseline result, and better compare the state of deep learning approaches. In our deep learning methodology, we have implemented several architectures (U-Net, LinkNet, and nnU-Net) along different hyperparameter tuning to achieve the best results. This process was achieved by:

%=====================================================================
\begin{figure}
\includegraphics[width=\textwidth]{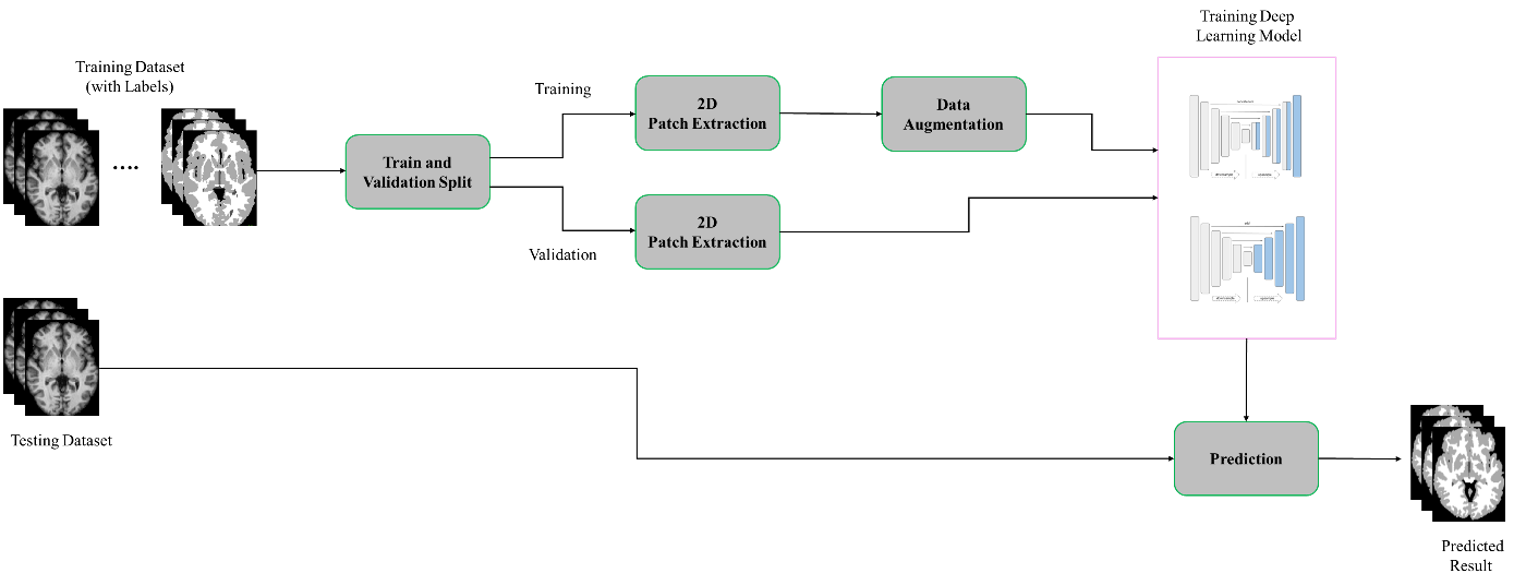}
\caption{ Pipeline for brain tissue segmentation using deep learning.
} \label{fig1}
\end{figure}
%=====================================================================

\begin{enumerate} 
\item  \textit{U-Net hyperparameter tuning:} We utilized the U-Net \cite{ronneberger2015u} architecture as our first deep learning segmentation model, however, we modified it to include a large number of feature channels inside the up-sampling section. These additional feature channels allow our network to propagate context information to higher-resolution layers. We then explored three U-Net architectures including customized models and models with pre-trained backbones to segment the brain tissues (CSF, GM, WM). The first model is the most straightforward, with single 2D Convolution layers in each step (encoding, bottleneck, and decoding) and standard up-sampling and skip connections. The second model increases the complexity of the architecture by adding two 2D Convolution layers in each step and incorporating connection operations that merge features to preserve spatial information. Finally, the third model is the most complex with the introduction of additional concatenation steps after each 2D Convolution layer to enhance the feature integration.

Apart from these three customized U-Net models, we also extended to U-Net architectures with pre-trained backbones using the large ImageNet dataset. Firstly, we used pre-trained ResNet34 and ResNet50 models as a backbone (encoding path) of the U-Net model for extracting features from the images. The decoder path remains the same as the conventional U-Net architecture, which utilizes up-sampling layers and convolutional processes to give the desired mask. This integration of a pre-trained backbone makes the U-Net architecture more complex and enhances the segmentation results.

%=====================================================================

\item  \textit{LinkNet hyperparameter tunning:} LinkNet \cite{chaurasia2017linknet} is a highly efficient deep learning model particularly used for real-time semantic segmentation due to its faster computational capacity. Just like other segmentation models, LinkNet architecture is comprised of two main parts: an encoder and a decoder. The encoder, which is based on the ResNet architecture, uses layers with 3x3 kernel sizes to capture high-level features while reducing dimensions. On the other hand, the decoder focuses on being lightweight and utilizes upsampling layers and 1x1 convolutions to reconstruct a detailed segmentation map. In this task, we used the pre-trained ResNet34 model as a backbone in the encoder part, similar to the step in our U-Net architecture hyperparameter tuning.

%=====================================================================
\begin{table}[h!]
\centering
\caption{Training Hyper-parameter for Different Deep Learning Models}
\begin{tabularx}{\textwidth}{|X|c|c|c|c|c|}
\hline
\textbf{Models} & \textbf{Epochs} & \textbf{Batch Size} & \textbf{Optimizer} & \textbf{Learning Rate} & \textbf{Loss Function} \\
\hline
2D U-Net-1 & 30 & 32, 64 & Adam & 0.0001 & Focal, Dice \\
2D U-Net-2 & 34 & 32, 64 & Adam & 0.0001 & Focal, Dice \\
2D U-Net-3 & 30 & 32, 64 & Adam & 0.0001 & Focal, Dice \\
ResNet34 U-Net & 32 & 64 & Adam & 0.0001 & Focal, Dice  \\
ResNet50 U-Net & 64 & 64 & Adam & 0.0001 & Focal, Dice  \\
2D nnU-Net & 300 & - & - & - & Dice  \\
3D nnU-Net & 115 & - & - & - & Dice  \\
ResNet34 LinkNet & 34 & 32, 64 & Adam & 0.0001 & Focal, Dice \\
\hline
\end{tabularx}
\end{table}

%=====================================================================

%=====================================================================

\item  \textit {nnU-Net:} nnU-Net \cite{isensee2021nnUNet} is widely used for semantic segmentation, especially in the biomedical imaging fields. The model has been designed to autonomously adapt and optimize itself by encompassing the customization of preprocessing methods, network architecture, training procedures, and post-processing techniques for any assigned new task. Therefore, less hyperparameter tunning is required to achieve the desired outcomes. Additionally, by default, it creates three distinct U-Net configurations such as a two-dimensional (2D) U-Net, a three-dimensional (3D) U-Net for operating the entire resolution of images, and a three-dimensional (3D) Cascade U-Net where the first U-Net works with downsampled images and the second is trained to enhance the full-resolution segmentation maps produced by the former.

For setting up the required environment for preparing datasets for the nnU-Net training and prediction, we created the directories and maintained the specific formats required for the nnU-Net model. After that, we prepared the dataset (in JSON format) by specifying the imaging modality (MRI) and labels (Background, CSF, GM, WM) for the training. Furthermore, we executed the pre-processing command to convert the dataset into the nnU-Net format. We then performed training indicating the desired configuration (2D/3D nnU-Net), and the resulting model was used to predict the unseen test dataset. Figure 4 shows the workflow of the deep learning steps while Table 1 shows several hyperparameters that were used during the training of the models.

\end{enumerate}

\section{Experiments and Results}

\subsection{Evaluation Metrics}
To evaluate the performances of our brain tissue segmentation models, we used the Dice Score (DSC), Hausdorff Distance (HD), and Absolute Volumetric Distance (AVD). The Dice Score is a similarity coefficient that measures the spatial overlap between the predicted and true segmentation masks, the Hausdorff Distance measures the maximum distance between any point in the predicted segmentation mask to the nearest point in the ground truth mask and vice versa, and finally, the Absolute Volumetric Distance measures the absolute difference in volume between the predicted and ground truth masks. They can be expressed mathematically as;

\vspace{0.2cm}
\begin{equation}
\text{Dice Score, DSC} = \frac{2 \times |A \cap B|}{|A| + |B|}
\end{equation}

\vspace{0.2cm}
\begin{equation}
\text{Hausdorff Distance, HD} (A, B) = \max \left\{ \sup_{a \in A} \inf_{b \in B} d(a, b), \sup_{b \in B} \inf_{a \in A} d(b, a) \right\}
\end{equation}

\vspace{0.2cm}
\begin{equation}
\text{Absolute Volumetric Distance, AVD} = \frac{|V_p - V_{gt}|}{V_{gt}} \times 100
\end{equation}

\vspace{0.2cm}
Where; \( A \) is the set of voxels in the predicted mask, and \( B \) is the set of voxels in the ground truth mask, \( |A \cap B| \) represents the number of common voxels between the predicted and ground truth masks, \( |A| \) and \( |B| \) are the total number of voxels in the predicted and ground truth masks, respectively. \( V_p \) is the volume of the predicted mask, \( V_{gt} \) is the volume of the ground truth mask, and \( |V_p - V_{gt}| \) is the absolute difference in volume.

%====================================================================

\subsection{Performance Analysis of Probabilistic ATLAS Based Brain Tissue Segmentation
}

In this comprehensive study, we used three different image registration parameters such as affine, rigid, and B-spline for performing the registration to assess how varying registration parameters affect segmentation performance. Following this, we computed key performance metrics, namely Dice Similarity Score (DSC), Hausdorff Distance (HD), and Absolute Volumetric Difference (AVD), the results of which are comprehensively presented in Table 2.

From the result in Table 2, the DSC results revealed a clear advantage of the affine registration technique over the rigid method, especially in the segmentation of CSF and GM. With a mean DSC value of 0.720 across different tissues, the affine method demonstrated its superior ability to adapt to brain structure variations, due to its scaling and shearing capabilities. This contrasted with the rigid method having limited rotation and translation capabilities yielded a moderate average DSC of 0.689. The B-spline technique achieved a lower average DSC of 0.561.

The affine method also outperformed the others using HD with the smallest average distance of about 6.349. The rigid method followed with a slightly higher average HD of 6.620 while the B-spline method showed the highest average HD of 7.002, suggesting that its adjustments might have introduced boundary distortions. The AVD metric presented more varied results as the rigid method exhibited a balanced average AVD of 14.806, while the affine method showed mixed results, with a notable variation in different tissues but better performance in WM with an AVD of 6.349. The B-spline method recorded the highest AVD values, averaging 28.283.

%===================================================================

\begin{table}[h!]
\centering
\caption{Performance Analysis of Probabilistic ATLAS Based Brain Tissue Segmentation}
\begin{tabular}{|c|c|c|c|c|}
\hline
\textbf{Parameter} & \textbf{CSF} & \textbf{GM} & \textbf{WM} & \textbf{Average Mean} \\
\hline
\multicolumn{5}{|c|}{\textbf{Mean Dice Coefficient Score (DSC)}} \\
\hline
Rigid & 0.624 $\pm$ 0.078 & 0.758 $\pm$ 0.036 & 0.688 $\pm$ 0.021 & 0.689 $\pm$ 0.046 \\
Affine & 0.674 $\pm$ 0.058 & 0.784 $\pm$ 0.037 & 0.704 $\pm$ 0.014 & 0.720 $\pm$ 0.036 \\
Affine + B-Spline & 0.454 $\pm$ 0.334 & 0.641 $\pm$ 0.196 & 0.587 $\pm$ 0.161 & 0.561 $\pm$ 0.230 \\
\hline
\multicolumn{5}{|c|}{\textbf{Mean Hausdorff Distance (HD)}} \\
\hline
Rigid & 4.416 $\pm$ 0.343 & 7.929 $\pm$ 1.186 & 7.517 $\pm$ 7.517 & 6.620 $\pm$ 0.623 \\
Affine & 4.298 $\pm$ 0.387 & 7.527 $\pm$ 0.414 & 7.222 $\pm$ 0.229 & 6.349 $\pm$ 0.343 \\
Affine + B-Spline & 4.749 $\pm$ 1.143 & 8.311 $\pm$ 0.948 & 7.947 $\pm$ 1.023 & 7.002 $\pm$ 1.038 \\
\hline
\multicolumn{5}{|c|}{\textbf{Mean Absolute Volumetric Difference (AVD)}} \\
\hline
Rigid & 17.244 $\pm$ 12.168 & 08.888 $\pm$ 06.720 & 18.286 $\pm$ 11.700 & 14.806 $\pm$ 10.196 \\
Affine & 19.312 $\pm$ 11.670 & 14.756 $\pm$ 12.516 & 11.002 $\pm$ 07.735 & 15.689 $\pm$ 10.454 \\
Affine + B-Spline & 36.005 $\pm$ 28.773 & 11.890 $\pm$ 10.745 & 36.947 $\pm$ 25.895 & 28.281 $\pm$ 21.806 \\
\hline
\end{tabular}
\end{table}
%====================================================================

\subsection{Performance Analysis of Deep Learning Based Brain Tissue Segmentation}

In our investigation using the deep learning models, we specifically assessed the performance of nnU-Net, LinkNet, U-Net, and the impact of incorporating backbones like ResNet34 and ResNet50. The performance analysis of the deep learning-based brain tissue segmentation is shown in Table 3. 

The nnU-Net, especially the 3D variant, exhibited exceptional performance across all tissues. The model achieved a DSC of 0.920 $\pm$ 0.018 for CSF, which is indicative of its precise segmentation capability. This performance was further confirmed with a DSC of 0.948 $\pm$ 0.007 for GM and 0.943 $\pm$ 0.012 for WM. The nnU-Net model also maintained superior boundary definition with the lowest HD values of 3.129 $\pm$ 0.451 for CSF, 6.192 $\pm$ 0.241 for GM, and 5.595 $\pm$ 0.359 for WM indicating a high level of accuracy in capturing the true margins of the brain tissues.

The ResNet50 U-Net, and ResNet34 U-Net models equally showed promising results. For example, the ResNet50 U-Net achieved a DSC of 0.910 $\pm$ 0.017 for GM, underscoring its robustness in segmentation despite the inherent complexity of neural structures while the ResNet34 U-Net demonstrated a solid performance with a mean DSC of 0.851 $\pm$ 0.040 across all tissues, and a mean HD of 5.985 $\pm$ 0.353, reflecting its effectiveness in maintaining boundary precision.

\vspace{0.5cm}
Additionally, the ResNet34 LinkNet variant achieved a DSC of 0.796 $\pm$ 0.047 for CSF, and a good mean HD value of 4.020 $\pm$ 0.206, suggesting a slight trade-off between the model's efficiency and the ultimate boundary accuracy compared to the nnU-Net. Also, the nnU-Net model demonstrated a strong volumetric consistency with an AVD of 5.766 $\pm$ 4.218 for CSF, outperforming the ResNet variants and LinkNet in volume estimation accuracy. Figures 5, 6, and 7 show the segmentation results of the best-performed model (3D nnU-Net), statistical analysis(3D nnU-Net), and the segmentation results from all models respectively.

%====================================================================

\begin{figure}
\includegraphics[width=\textwidth]{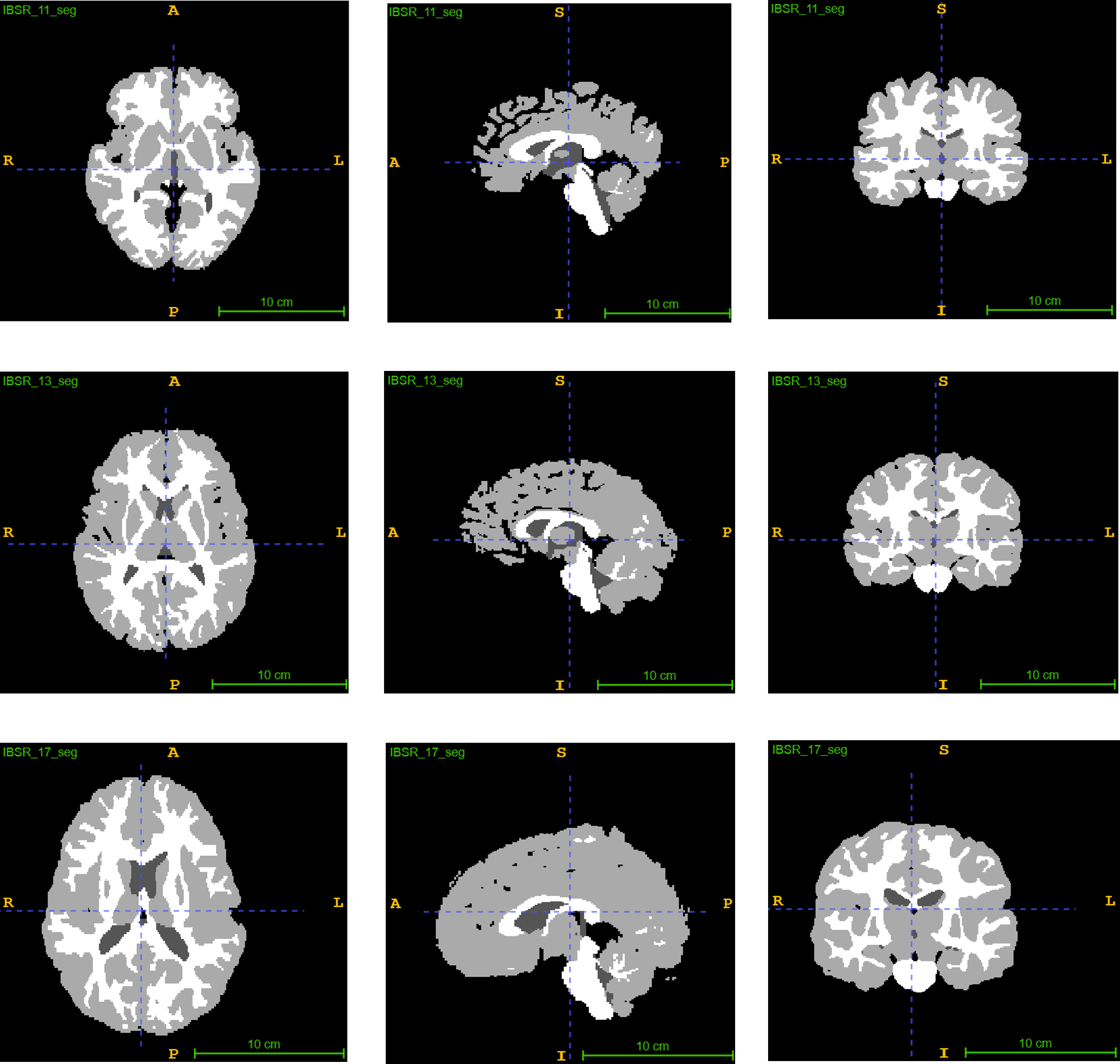}
\caption{ Segmentation results of the best model (3D nnU-Net) with three different views (axial, sagittal, coronal).
} \label{fig1}
\end{figure}
%====================================================================

\begin{figure}
\includegraphics[width=\textwidth]{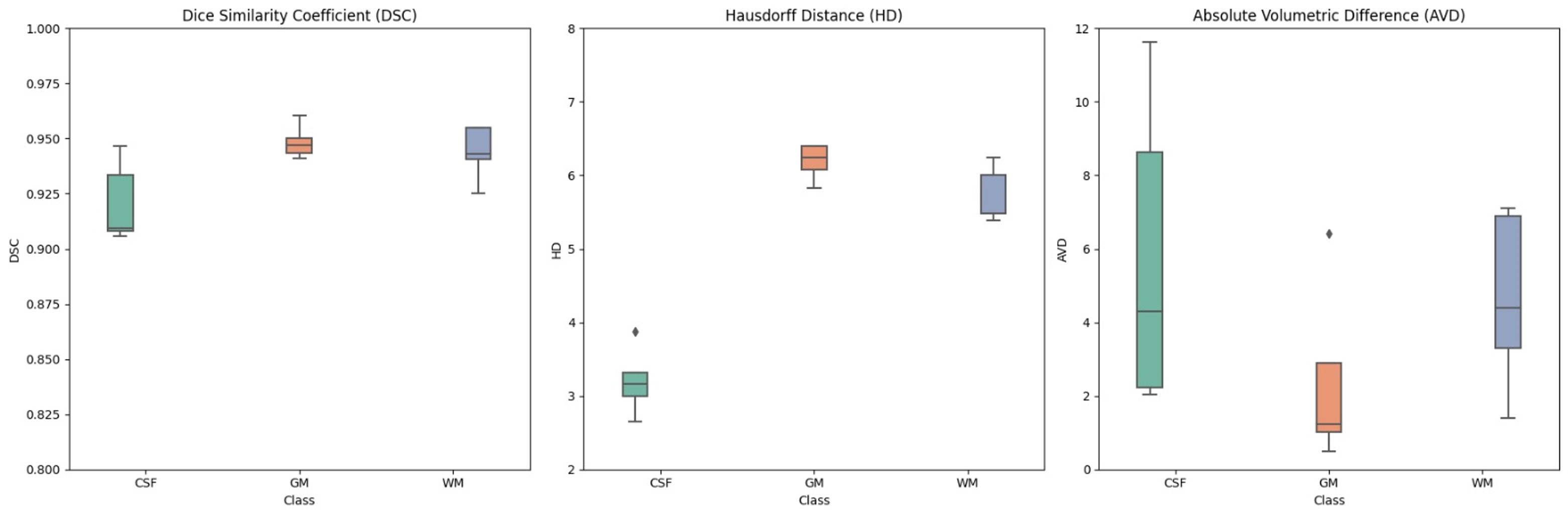}
\caption{Statistical analysis of the best-performed mode (3D nnU-Net)
} \label{fig1}
\end{figure}
%====================================================================
\begin{figure}
\includegraphics[width=\textwidth]{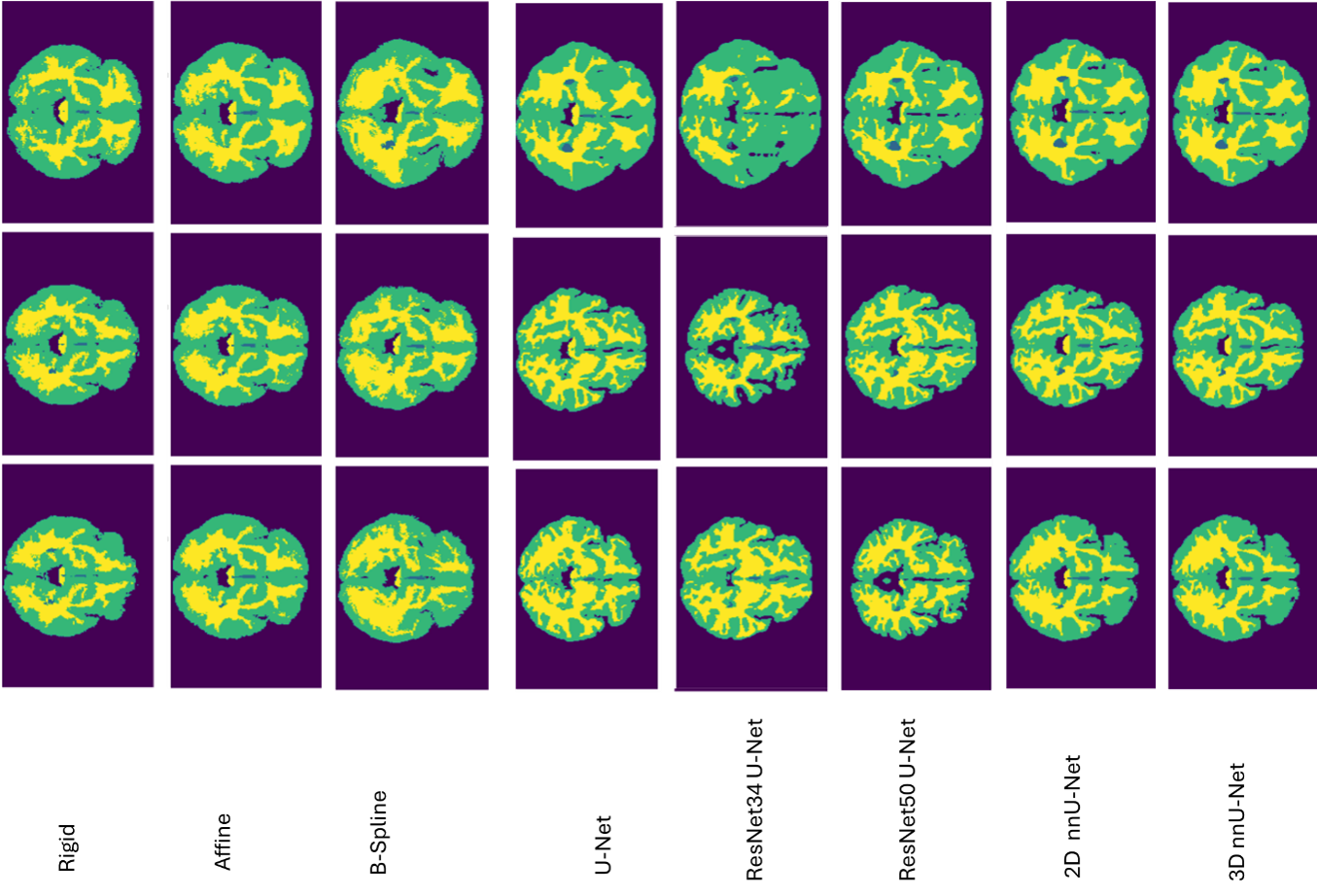}
\caption{Segmentation Results from all models.
} \label{fig1}
\end{figure}
%====================================================================

\begin{table}[h!]
\centering
\caption{Performance Analysis of Deep Learning Based Brain Tissue Segmentation}
\begin{tabular}{|c|c|c|c|c|}
\hline
\textbf{Method} & \textbf{CSF} & \textbf{GM} & \textbf{WM} & \textbf{Average Mean} \\
\hline
\multicolumn{5}{|c|}{\textbf{Mean Dice Coefficient Score (DSC)}} \\
\hline
2D U-Net-1 & 0.659 $\pm$ 0.082 & 0.884 $\pm$ 0.015 & 0.838 $\pm$ 0.071 & 0.793 $\pm$ 0.005 \\
2D U-Net-2 & 0.759 $\pm$ 0.062 & 0.892 $\pm$ 0.018 & 0.835 $\pm$ 0.084 & 0.829 $\pm$ 0.055 \\
2D U-Net-3 & 0.736 $\pm$ 0.107 & 0.898 $\pm$ 0.013 & 0.849 $\pm$ 0.046 & 0.828 $\pm$ 0.049 \\
ResNet34 U-Net & 0.661 $\pm$ 0.104 & 0.814 $\pm$ 0.053 & 0.737 $\pm$ 0.135 & 0.737 $\pm$ 0.105 \\
ResNet50 U-Net & 0.814 $\pm$ 0.018 & 0.910 $\pm$ 0.017 & 0.908 $\pm$ 0.017 & 0.877 $\pm$ 0.007 \\
2D nnU-Net & 0.914 $\pm$ 0.018 & 0.948 $\pm$ 0.007 & 0.943 $\pm$ 0.012 & 0.937 $\pm$ 0.012 \\
3D nnU-Net & 0.920 $\pm$ 0.011 & 0.948 $\pm$ 0.007 & 0.948 $\pm$ 0.006 & 0.938 $\pm$ 0.009 \\
ResNet34 LinkNet & 0.796 $\pm$ 0.047 & 0.894 $\pm$ 0.020 & 0.862 $\pm$ 0.054 & 0.851 $\pm$ 0.040 \\
\hline
\multicolumn{5}{|c|}{\textbf{Mean Hausdorff Distance (HD)}} \\
\hline
2D U-Net-1 & 5.510 $\pm$ 0.569 & 7.296 $\pm$ 0.457 & 6.296 $\pm$ 0.724 & 6.243 $\pm$ 0.583 \\
2D U-Net-2 & 4.016 $\pm$ 0.289 & 7.220 $\pm$ 0.579 & 6.811 $\pm$ 0.712 & 6.016 $\pm$ 0.528 \\
2D U-Net-3 & 4.309 $\pm$ 0.173 & 7.334 $\pm$ 0.500 & 6.874 $\pm$ 0.826 & 6.172 $\pm$ 0.500 \\
ResNet34 U-Net & 4.429 $\pm$ 0.169 & 7.050 $\pm$ 0.630 & 7.994 $\pm$ 1.119 & 6.491 $\pm$ 0.771 \\
ResNet50 U-Net & 4.079 $\pm$ 0.486 & 6.195 $\pm$ 0.440 & 5.560 $\pm$ 0.440 & 5.278 $\pm$ 0.443 \\
2D nnU-Net & 3.226 $\pm$ 0.486 & 6.193 $\pm$ 0.210 & 5.595 $\pm$ 0.329 & 5.005 $\pm$ 0.343 \\
3D nnU-Net & 3.129 $\pm$ 0.451 & 6.192 $\pm$ 0.241 & 5.595 $\pm$ 0.349 & 4.972 $\pm$ 0.340 \\
ResNet34 LinkNet & 4.020 $\pm$ 0.206 & 7.280 $\pm$ 0.496 & 6.655 $\pm$ 0.356 & 5.985 $\pm$ 0.353 \\
\hline
\multicolumn{5}{|c|}{\textbf{Mean Absolute Volumetric Difference (AVD)}} \\
\hline
2D U-Net-1 & 47.893 $\pm$ 62.352 & 06.749 $\pm$ 06.003 & 15.388 $\pm$ 24.260 & 23.343 $\pm$ 30.872 \\
2D U-Net-2 & 33.524 $\pm$ 26.936 & 08.719 $\pm$ 05.268 & 14.049 $\pm$ 19.252 & 18.764 $\pm$ 17.152 \\
2D U-Net-3 & 20.297 $\pm$ 19.395 & 13.235 $\pm$ 10.570 & 12.962 $\pm$ 11.816 & 15.498 $\pm$ 13.260 \\
ResNet34 U-Net & 28.914 $\pm$ 30.845 & 09.056 $\pm$ 09.056 & 19.134 $\pm$ 20.382 & 19.035 $\pm$ 20.094 \\
ResNet50 U-Net & 21.974 $\pm$ 19.634 & 08.749 $\pm$ 07.549 & 09.518 $\pm$ 08.064 & 13.414 $\pm$ 11.082 \\
2D nnU-Net & 09.402 $\pm$ 11.408 & 09.092 $\pm$ 08.056 & 09.622 $\pm$ 07.586 & 09.372 $\pm$ 09.020 \\
3D nnU-Net & 09.284 $\pm$ 10.508 & 09.092 $\pm$ 08.002 & 09.582 $\pm$ 07.632 & 09.319 $\pm$ 08.714 \\
ResNet34 LinkNet & 29.002 $\pm$ 14.329 & 09.092 $\pm$ 07.652 & 09.822 $\pm$ 08.062 & 15.972 $\pm$ 10.681 \\
\hline
\end{tabular}
\end{table}
%====================================================================

\section{Discussion}

The results from the probabilistic atlas show that while the affine transformation offers the most balanced approach in terms of DSC and HD, it does not uniformly excel in AVD, highlighting the decision-making required in choosing registration techniques. The rigid method, despite its simplicity, provides a robust baseline in some scenarios. The B-spline technique, despite its potential for precise local adjustments, does not consistently translate into better segmentation performance, potentially due to overfitting issues where the B-spline transformation over-adjusts to the atlas. This analysis underscores the importance of carefully selecting registration parameters in atlas-based segmentation, considering both the specific characteristics of brain tissues and the desired outcomes of the segmentation process. 

For the deep learning techniques, the comprehensive analysis presented reveals that the 3D nnU-Net stands out as a clear leader in terms of dice score, boundary precision, and volumetric consistency. The ResNet-infused U-Nets also show commendable performance, especially in the segmentation of grey matter. The LinkNet architecture, while slightly less precise in boundary definition, offers a balance between efficiency and performance, potentially useful in scenarios where computational resources or time are limiting factors. These insights reflect the nuanced trade-offs present in selecting a deep learning model for brain tissue segmentation and highlight the substantial progress that has been made in the application of these advanced computational techniques to medical imaging.

The segmented images, as shown in Figure 7, qualitatively confirm the proficiency of all models, with the 3D nnU-Net's results being distinctly superior. The visual clarity and precision of the 3D nnU-Net model's segmentation are further illustrated through images segmented across the three anatomical planes: axial, coronal, and sagittal, as depicted in Figure 5. This evidence supports the conclusion that, when deep learning approaches are implemented with appropriate hyperparameters, they can outperform statistical methods in medical image segmentation.

\section{Conclusion}

In conclusion, this study demonstrates that advanced deep learning models, particularly the 3D nnU-Net, significantly outperform traditional probabilistic atlas-based methods in brain tissue segmentation from MRI. The 3D nnU-Net excels in dice score, boundary precision, and volumetric consistency, highlighting its potential for high-quality segmentation in medical imaging.

For future work, further exploration into refining the 3D nnU-Net model and optimizing transformer-based models is recommended to achieve even greater efficiency. Additionally, investigating the integration of these models into clinical workflows could expedite and enhance the diagnostic process, potentially improving patient outcomes. Expanding the dataset to encompass a broader range of imaging conditions and diverse patient populations would also be beneficial in developing a more universally applicable model.

%====================================================================
\begin{credits}
\subsubsection{\ackname} We are grateful for the skills and knowledge gained from the MSc in Medical Imaging and Applications (MAIA) program that enabled us to work on this innovative project.

\subsubsection{\discintname}
The authors have no competing interests to declare that are
relevant to the content of this article.
\end{credits}
%
% ---- Bibliography ----
%
% BibTeX users should specify bibliography style 'splncs04'.
% References will then be sorted and formatted in the correct style.
%
%\bibliographystyle{splncs04}
\bibliographystyle{IEEEtran}
\bibliography{references}

\end{document}